\documentclass[aps,pra,twocolumn,showpacs]{revtex4-2}
\usepackage{graphicx}
\usepackage{amssymb,amsfonts,amsmath}
\usepackage[colorlinks=true,citecolor=Cerulean,linkcolor=RubineRed,urlcolor=Cerulean]{hyperref}

\hypersetup{breaklinks=true}
\usepackage{graphicx}
\usepackage{color}
\usepackage[usenames,dvipsnames]{xcolor}
\usepackage{epstopdf}
\usepackage[normalem]{ulem}
\usepackage{bm}
\usepackage{bbold}

\usepackage[american]{babel}
\usepackage{lipsum}
\usepackage{hyperref}
\hypersetup{
    colorlinks=true,
    linkcolor=magenta,
    filecolor=magenta,      
    urlcolor=cyan,
    pdftitle={Overleaf Example},
    pdfpagemode=FullScreen,
    }

\renewcommand{\d}{\mathrm{d}}

\renewcommand{\i}{{\rm i}}

\renewcommand{\vec}[1]{{\bf #1}}

\begin{document}

\title{
Efficient Microwave Spin Control of Negatively Charged Group-IV Color Centers in Diamond
}

\author{Gregor Pieplow}
\altaffiliation{These authors contributed equally to this work}
\affiliation{Department of Physics, Humboldt-Universit\"{a}t zu Berlin, Newtonstr. 15, 12489 Berlin, Germany}

\author{Mohamed Belhassen}
\altaffiliation{These authors contributed equally to this work}
\affiliation{Department of Physics, Humboldt-Universit\"{a}t zu Berlin, Newtonstr. 15, 12489 Berlin, Germany}

\author{Tim Schr\"{o}der}
\email[Corresponding author: ]{tim.schroeder@physik.hu-berlin.de}
\affiliation{Department of Physics, Humboldt-Universit\"{a}t zu Berlin, Newtonstr. 15, 12489 Berlin, Germany}
\affiliation{Ferdinand-Braun-Institut, Gustav-Kirchhoff-Straße 4, 12489 Berlin, Germany}

\date{\today}

\begin{abstract}
In this work, we provide a comprehensive overview of the microwave-induced manipulation of electronic spin states in negatively charged group-IV color centers in diamond with a particular emphasis on the influence of strain. Central to our investigation is the consideration of the full vectorial attributes of the magnetic fields involved, which are a \textit{dc} field for lifting the degeneracy of the spin levels and an \textit{ac} field for microwave control between two spin levels. We observe an intricate interdependence between their spatial orientations, the externally applied strain, and the resultant efficacy in spin state control. In most work to date the \textit{ac} and \textit{dc} magnetic field orientations have been insufficiently addressed, which has led to the conclusion that strain is indispensable for the effective microwave control of heavier group-IV vacancies, such as tin- and lead-vacancy color centers.
In contrast, we find that the alignment of the \textit{dc} magnetic field orthogonal to the symmetry axis and the \textit{ac} field parallel to it can make the application of strain obsolete for effective spin manipulation. Furthermore, we explore the implications of this field configuration on the spin's optical initialization, readout, and gate fidelities.
\end{abstract}
\maketitle

\section{Introduction}

    For many quantum information applications \cite{childress_diamond_2013} such as quantum sensing \cite{degen_quantum_2017}, quantum networks \cite{nguyen_integrated_2019, bhaskar_experimental_2020} and quantum computing \cite{gambetta_building_2017}, 
    key requirements include ultra-high fidelity initialization, both single and two-qubit gate operations, as well as readout. Notably, the execution of quantum algorithms that offer a  quantum advantage \cite{shor_algorithms_1994} demand near-perfect operations within the coherence time of the system. In particular, the fidelity of single-qubit gates now consistently exceeds 99\% across diverse platforms, including superconducting qubits \cite{wang_experimental_2018}, quantum dots \cite{yoneda_quantum-dot_2018}, trapped ions \cite{harty_high-fidelity_2014}, and ultra-cold atoms \cite{xia_randomized_2015}. 
    
    An emerging class of optically active spin-1/2 systems for quantum information applications is found in negatively charged group-IV color centers in diamond (G4Vs) \cite{thiering_ab_2018}, including silicon (SiV), germanium (GeV), tin (SnV), and lead vacancy center (PbV). G4Vs exhibit notably high optical efficiency, with a Debye-Waller factor of 60-80\% \cite{neu_single_2011,palyanov_germanium_2015,gorlitz_spectroscopic_2020}), in contrast to the 3\% Debye-Waller factor observed in the negatively charged nitrogen vacancy center \cite{santori_nanophotonics_2010}. Moreover, they display lower sensitivity to electric noise \cite{sipahigil_indistinguishable_2014}, rendering them highly suitable for integration into nanostructures \cite{nguyen_integrated_2019, rugar_narrow-linewidth_2020}, which is crucial for efficient coupling to a precisely defined optical mode \cite{bopp_sawfish_2022}. Furthermore, G4Vs exhibit excellent spin coherence times, with, for example, the GeV showing $T_2 = 20$ ms \cite{senkalla_germanium_2023}.
    \begin{figure}[t!]
    \centering
        \includegraphics[width=.9\columnwidth]{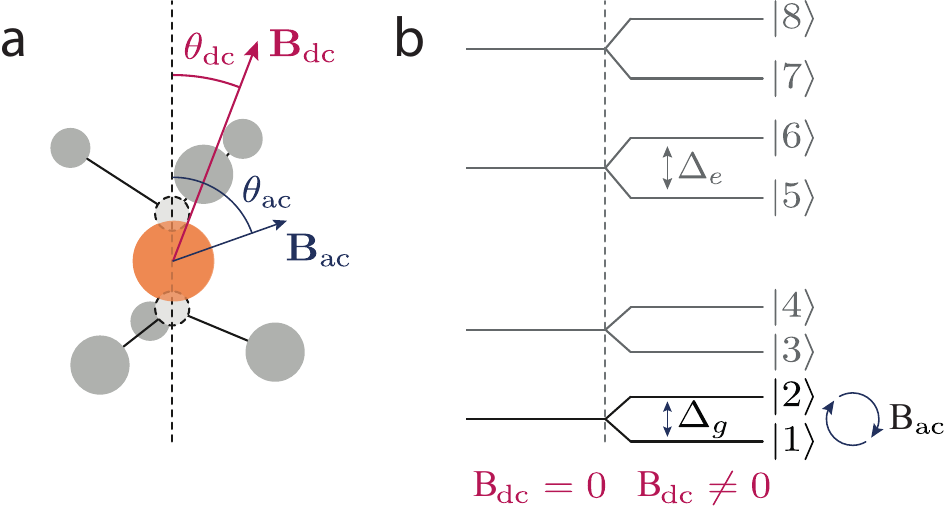}
        \caption{Illustration of the physical system.
        a) group-IV color center with the respective angles of the \textit{dc}
        magnetic field $\vec{B}_{\rm dc}$ and the \textit{ac} driving field $\vec B_{\rm ac}$ relative to the symmetry axis. 
        b) Level scheme of the ground and excited states of the color center (for details see section \ref{section:migates}). For $B_{\rm dc}=0$, the ground and excited states are spin-degenerate. The \textit{dc} magnetic field lifts the spin degeneracy. The qubit driven by $\vec B_{\rm ac}$ is
        composed of the states $|1\rangle$ and $|2\rangle$ with a level splitting of $\Delta_g$. $\Delta_e$ is the splitting between the first two excited state levels.
        }
        \label{fig:fig0}
    \end{figure}

    This work primarily focuses on the theoretic analysis of single-qubit gates acting on the electronic spin, which operate in the microwave regime. In the first works demonstrating coherent microwave control of the SnV \cite{rosenthal_microwave_2023, guo_microwave-based_2023} a consensus has emerged, which asserts that strain is an essential prerequisite for manipulating the spin of the heavier implanted group-IV elements such as tin and lead, owing to their increased spin-orbit coupling in comparison to lighter elements like silicon and germanium. Consequently, we are particularly interested in the interplay of strain and spin-orbit coupling in the investigation of coherent control of the spin qubit. 
    
    Interestingly, we discovered a magnetic field configuration, for which strain is not only unnecessary but actually hampers the efficiency of the coherent control. As we will demonstrate that the efficiency depends on the orientation of the static field $\vec{B}_{\rm dc}$, which is necessary to lift the spin degeneracy, and the oscillating field $\vec{B}_{\rm ac}(t)$, responsible for driving the electronic spin (see Fig.~\ref{fig:fig0}a).\\
  
    This work is structured as follows: We first provide a brief overview of the static contributions to the systems Hamiltonian, such as the spin-orbit coupling, the Jan-Teller effect, the interaction with internal and external strain as well as the static magnetic field $\vec B_{\rm dc}$. We then introduce the influence of $\vec{B}_{\rm ac}(t)$ and explain how such a field generates single-qubit gates acting on the two energetically lowest lying qubit states $|1\rangle$ and $|2\rangle$ (see Fig.\ref{fig:fig0}b).    
    
    We find a closed approximate expression for the qubit's Rabi frequency $\Omega$ as a function of total strain acting on the G4V, the spin orbit coupling and the magnetic fields $\vec{B}_{\rm dc}$ and $\vec{B}_{\rm ac}$. 
    The Rabi frequency $\Omega$ is a key performance metric for the driving efficiency: Increasing $\Omega$ independently from the magnetic field strength $B_{\rm ac}$ renders the magnetic control more efficient as it reduces the need for microwave power, resulting in lower levels of heating and, in broader terms, a decrease in the experimental overhead needed for spin control. Notably, for the most efficient choice of control parameters, i.e. for perpendicular $\vec{B}_{\rm dc}$, strain reduces the efficiency independent of the G4V as can be seen in, for example, Figs.~\ref{fig:Rabi_Frequency Strain Strain}c and d, where the Rabi frequency decreases, for both the SiV and SnV for increasing values of strain.
    
    Finally we analyse initialization and readout based on the master equations in Lindblad form for various configurations of the magnetic field, and find that for the most efficient control regime, both are possible.
  
\section{Microwave gates}
\label{section:migates}
    \begin{figure}[t!]
    \includegraphics[width=.9\columnwidth]{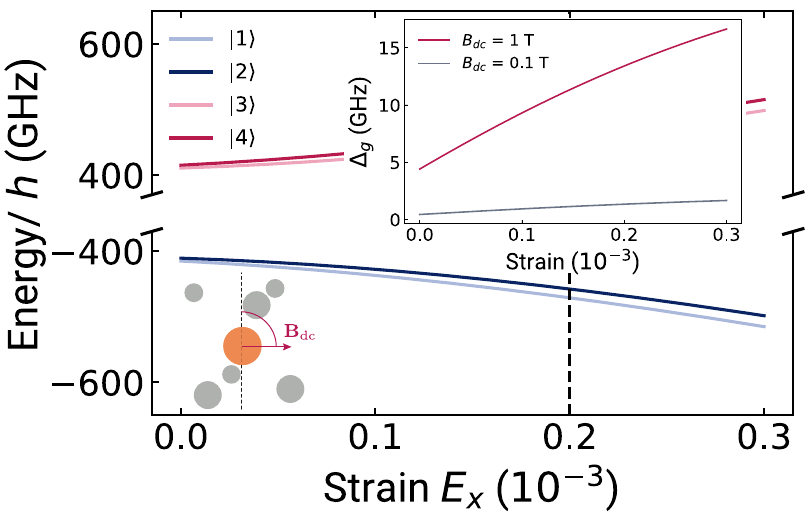}
    \caption{
    The relative energies of the ground state orbital branches $|1\rangle, |2\rangle$ and $|3\rangle, |4\rangle$ of the SnV as a function of strain in the presence of $B_{\rm dc} = 1$ T oriented perpendicular to the symmetry axis. The value of 1 T is used to enhance the visibility of the splitting. The inset shows the splitting $\Delta_g$ between $|1\rangle$ and $|2\rangle$ for $B_{\rm dc} = 0.1$ T and 1 T. The dashed vertical line marks a comparatively large but experimentally realistic strain value \cite{meesala_strain_2018}. For $B_{\rm dc} > 0.1$\,T the ground state splitting $\Delta_g > 1/\tau_{\rm rad}$, where $\tau_{\rm rad} = 4.5$ ns is the radiative lifetime of the SnV \cite{trusheim_transform-limited_2020}.  
    }
    \label{fig:fig0c}
    \end{figure}

    We decompose the total magnetic field at the position of the SnV into a static \textit{dc} and time-dependent \textit{ac} part $\vec{B}(t) = \vec{B}_{\rm dc} + \vec{B}_{\rm ac}(t)$, where $\vec B_{\rm ac}$ is either the near-field of an alternating current passing through a stripline in close proximity to the defect, or the far field of a microwave source. In this work, irrespective of the generation $\vec B_{\rm ac}(t)$ is referred to as the microwave drive of the system.  
    
    Based on the decomposition of the magnetic field we split the Hamiltonian generating the time evolution of the system into a \textit{dc} and \textit{ac} component
    \begin{equation}
        H(t) = H_{\rm dc} + H_{\rm ac}(t)~, 
        \label{eq:time_dep_hamiltonian}
    \end{equation}
    where $H_{\rm dc}$ generates the free evolution and $H_{\rm ac}$ is the interaction of the system with $B_{\rm ac} (t)$. We assume that $B_{\rm ac} (t)$ drives the system for some finite gate time $T_{\rm G}$. The single-qubit gate operating on the electronic spin is then given by 
    \begin{equation}
        U(t_e, t_i) = T e^{\frac{\i}{\hbar} \int_{t_i}^{t_e} H(t) \d t }~,
        \label{eq:gate}
    \end{equation}
    where $T_{\rm G} = t_e - t_i$ and $T$ is the time-ordering operator. We will confirm that indeed any desired single qubit gate $U\in$ SU(2) can be constructed with a suitable series of control pulses associated to $H(t)$.   
    In the following we detail the interactions contributing to $H_{\rm dc}$ and $H_{\rm ac}$. 
    We separate $H_{\rm dc}$ into
    \begin{equation}
        H_{\rm dc} = H_{\rm int.} + H_{\rm ext.}   
        \label{eq:ham_dc}
    \end{equation}
    Where $H_{\rm int.}$ contains all the terms that are intrinsic to the diamond system and 
    $H_{\rm ext.}$ contains terms such as 
    external strain or the static field $\vec{B}_{\rm dc}$. 
    
    The intrinsic components are
    \begin{equation}
        H_{\rm int.} = H_0 + H_{\rm SO} + H_{\rm JT} + H_{\rm int.~strain}~,
    \end{equation}
    where $H_0$ describes the unperturbed effective single particle system. The general structure of all contributions to $H_{\rm dc/ac}$ can be deduced from group theory and the D$_{\rm 3d}$ symmetry of the unperturbed system associated to G4V defects \cite{hepp_electronic}. $H_{\rm SO}$ is the spin orbit interaction and $H_{\rm JT}$ describes the Jahn-Teller effect. 
    Finally, $H_{\rm int.~strain}$ is the internal strain, by which we mean distortions of the carbon lattice due to the implanted ion. It is often combined with the Jahn-Teller contribution. The intrinsic components of the Hamiltonian are identical for all color centers of the same kind. 

    The extrinsic components to $H_{\rm dc}$ are
    \begin{equation}
        H_{\rm ext.} = H_{\rm ext.~strain} + H_{\rm B_{\rm dc}}  
    \end{equation}
    The \textit{dc} component of the magnetic field lifts the spin degeneracy through the Zeeman term and therefore makes the spin degree of freedom accessible to microwave control. External strain described by $H_{\rm ext.~strain}$ can have many causes, such as the displacement of lattice atoms by interstitial crystal impurities \cite{sumiya_distribution_2023}, lattice vacancies such as mono-, di- or multi-vacancy complexes \cite{fisher_vacancy_2006}, lattice dislocations \cite{willems_strain_2005} or it can be applied externally \cite{meesala_strain_2018}.
    
    In the following we describe the explicit structure of the interaction terms. Following \cite{hepp_electronic} a basis for an explicit representation of the interactions can be chosen using group theoretic arguments based on the D$_{\rm 3d}$ symmetry. For the discussion of the microwave control we focus on the ground state manifold, which is spanned by the four states
    \begin{equation}
        |e_{gx}\uparrow\rangle,~|e_{gx}\downarrow\rangle,~|e_{gy}\uparrow\rangle,~|e_{gy}\downarrow\rangle~,
        \label{eq:basis_states}
    \end{equation}
    which are energetically degenerate eigenstates of $H_0$ belonging to the $E$ irreducible representation. The two 
    orbital states labeled with $e_{gx/gy}$ have even symmetry and are two fold spin degenerate. Due to the degeneracy we can drop the $H_0$ contribution to the Hamiltonian and focus on the representation of the other internal and external contributions to the Hamiltonian. From \cite{hepp_electronic} the representations of the various contributions are 
    in the above basis \eqref{eq:basis_states}:
    \begin{equation}
    H_{\rm SO}=\begin{pmatrix}
    0 & -i \lambda\\
    i \lambda & 0 \\
    \end{pmatrix} \otimes \begin{pmatrix}
    1 & 0\\
    0 & -1 \\
    \end{pmatrix}
    \end{equation}
    where $\lambda$ is the spin-orbital coupling. Note that compared to \cite{hepp_electronic} we absorbed the $1/2$ into $\lambda$.
    \begin{equation}
    H_{\rm JT}= \begin{pmatrix}
    \Upsilon_x & \Upsilon_y\\
    \Upsilon_y & -\Upsilon_x \\
    \end{pmatrix} \otimes \begin{pmatrix}
    1 & 0\\
    0 & 1 \\
    \end{pmatrix}
    \end{equation}
    where $\Upsilon_x$ and $\Upsilon_y$ are the Jahn-Teller coupling strengths.
    
    The external and internal strain contributions are:
    \begin{equation}
    \label{eq:Strain}
    H_{\rm S}= \begin{pmatrix}
    dE_x & 2d\epsilon_{xy}\\
    2d\epsilon_{xy} & -dE_x \\
    \end{pmatrix} \otimes \begin{pmatrix}
    1 & 0\\
    0 & 1 \\
    \end{pmatrix}
    \end{equation}
    where 
    \begin{equation}
    E_x=\epsilon_{xx}-\epsilon_{yy}~,
    \end{equation}
    $\epsilon_{xx}$, $\epsilon_{yy}$ and $\epsilon_{xy}$ are the strain tensor elements representing uni-axial stress and $d$ is the spin-strain coupling strength. For $H_{\rm int.~strain}$ and $H_{\rm ext.~strain}$ we discriminate between $E_x^{\rm int.}, \epsilon_{\rm xy}^{\rm int.}$ and $E_x^{\rm ext.}, \epsilon_{\rm xy}^{\rm ext.}$. For the sake of brevity, we drop the superscript from the external contribution. The degeneracy of the four basis states \eqref{eq:basis_states} is lifted due to  $H_{\rm SO}$, $H_{\rm JT}$ and $H_{\rm int.~strain}$ resulting in two spin-degenerate levels as shown in Fig.\ref{fig:fig0}b for $B_{\rm dc} = 0$.       

    The static magnetic field $B_{\rm dc}$ lifts the spin degeneracy, which is required to turn the long lived spin states into a qubit that is amenable to microwave control. Magnetic fields couple to the vacancy according to \cite{hepp_electronic}
    \begin{align}
    H_{\rm B} &= H_{\parallel} + H_{\perp} \nonumber\\
        &\begin{aligned}
        = f&\gamma_L B_z\begin{pmatrix}
    0 & i \\
    -i & 0
    \end{pmatrix} \otimes \begin{pmatrix}
    1 & 0 \\
    0 & 1 \\
    \end{pmatrix} 
    \\+ &\gamma_S \begin{pmatrix}
    1 & 0 \\
    0 & 1 \\
    \end{pmatrix} \otimes  \begin{pmatrix}
    B_z & B_x - \i B_y \\
    B_x+ \i B_y & -B_z \\
    \end{pmatrix}
    \end{aligned}
    \label{eq:ham_magnetic}
    \end{align}    
    where the magnetic field is $\vec{B} = (B_x, B_y, B_z)$, and the $z$-direction is aligned with defect's symmetry axis. Furthermore,  $f = 0.15$ is a coupling constant and  
    \begin{equation}
        \gamma_L=\gamma_S=\frac{\mu_B}{\hbar}~,
    \end{equation}
    where $\mu_B$ is the Bohr magneton. In general $\gamma_L\neq\gamma_S$ due to the difference between the gyromagnetic ratios, however, here $\gamma_L = \gamma_S$ due to the factor 1/2 in the spin-1/2 Pauli operators. Both the $\vec{B}_{\rm dc}$ and $\vec{B}_{\rm ac}(t)$ fields couple to the system through Eq.~\eqref{eq:ham_magnetic}.
    \begin{table}[h!]
    \centering
    \begin{tabular}{c c c c c c} 
      & $2\lambda$ (GHz) & $\Upsilon_x$ (GHz)& $\Upsilon_y$ (GHz)& f & $d$ (PHz/Strain) \\ 
    \hline
    \hline
    \multicolumn{6}{c}{Ground States} \\
    \hline
     SiV & 49 & 2 & 3 & 0.10 & 1.3 \\
     GeV & 207 & NA & NA & NA & NA \\
     SnV & 815 & 65 & 0 & 0.15 & 0.787 \\
     PbV & 4385 & NA & NA & NA & NA
     \\
     \hline
    \multicolumn{6}{c}{Excited States} \\
    \hline
     SiV & 257 & 12 & 16 & 0.10 & 1.8 \\ 
     GeV & 989 & NA & NA & NA & NA \\
     SnV & 2355 & 855 & 0 & 0.15 & 0.956 \\
     PbV & 6920 & NA & NA & NA & NA\\[1ex] 
     \hline
     \hline
    \end{tabular}
    \caption{Parameters used in this paper for the unstrained SiV \cite{hepp_electronic_2014-1, meesala_strain_2018}, GeV \cite{thiering_ab_2018}, SnV \cite{trusheim_transform-limited_2020, guo_microwave-based_2023} and PbV \cite{thiering_ab_2018}. $\lambda$ is the spin-orbital coupling, $\Upsilon_x$ and $\Upsilon_y$ are the Jahn-Teller coupling strengths, $f$ is the quenching factor of $\gamma_L$, and $d$ is the spin-strain coupling constant. The SiV and GeV are categorized as light color centers, while the SnV and PbV are designated as heavy color centers.}
    \label{table:SiV SnV parameters}
    \end{table}
    In Fig.~\ref{fig:fig0}a, we illustrate the field orientations with respect to the defect's symmetry axis. In Fig.~\ref{fig:fig0}b we show the G4V energy levels including the excited states. The eigenstates of $H_{\rm dc}$ are then comprised of the four ground states ($|1\rangle$-$|4\rangle$). For the sake of completeness, we also included the four excited states ($|4\rangle$-$|8\rangle$). Structurally, the Jahn-Teller interaction and strain Hamiltonian are identical, which makes it impossible to separate the Jahn-Teller effect from internal strain experimentally. For that reason in experiments $H_{\rm int.~strain}$ is always combined with the contribution of $H_{\rm JT}$. In our theoretical analysis, we refer to zero strain when $H_{\rm ext.~strain} = 0$.

    As an example, we show in Fig.~\ref{fig:fig0c} the energy levels of the SnV's ground states as a function of strain at $B_{\rm dc} = 1$T (we use $1$ T to enhance the visibility of behaviors), orthogonal to the symmetry axis. Strain increases both the splitting between the orbital branches $|1\rangle, |2\rangle$ and $|3\rangle, |4\rangle$, which can lead to increased spin coherence times as thermal decoherence channels are reduced \cite{jahnke_electronphonon_2015}. The inset shows the splitting $\Delta$ between $|1\rangle$ and $|2\rangle$, which is $\propto B_{\rm dc}$. In Table~\ref{table:SiV SnV parameters} we present established model parameters for the light G4V (SiV and GeV) and the heavy G4V (SnV and PbV).
    
    Having set up a full model of the physical systems under consideration, we can now analyze the interplay of strain and the implementation of microwave enabled gates acting on the spin degree of freedom. For that purpose we begin with diagonalizing $H_{\rm dc}$ and writing $H_{\rm B_{\rm ac}(t)}$ in terms of the eigenstates of $H_{\rm dc}$ (see appendix \ref{app:effective}). The diagonalization of $H_{\rm dc}$ is done in two steps:
    First we diagonalize $H = H_{\rm int}+ H_{\rm ext. strain}$, for which we find two two-fold spin degenerate eigenstates with energies
    \begin{equation}
        e_{\pm, \downarrow\uparrow} = \pm\sqrt{\lambda^2 + \xi^2}~,
    \end{equation}
    where $\xi^2 = \upsilon_x^2 + \upsilon_y^2$, $\upsilon_x = (d E_x + \Upsilon_x)$, $\upsilon_y = (2 d \epsilon_{xy} + \Upsilon_y)$, $E_x = E_x^{\rm int} + E_x^{\rm ext}$ and $\epsilon_{xy} = \epsilon_{xy}^{\rm int.} + \epsilon_{xy}^{\rm ext.}$. We then re-express $H_{\rm B_{\rm dc}}$ in the energy eigenbasis  $|e_-, \downarrow\rangle, |e_-, \uparrow\rangle,$ $|e_+, \downarrow\rangle, |e_+, \uparrow\rangle$. The detailed expressions can be found in the appendix \ref{app:effective}. We can remove the states $|e_+, \uparrow\downarrow\rangle$ through adiabatic elimination as long as $2 e_{+,\downarrow}$ is bigger than any other energy in the problem \cite{paulisch_beyond_2013}. This reduction places certain limits on the magnetic field strengths and strain regimes for which the approximation is valid.     
    
    The reduction of the system results in 
    \begin{equation}
        H_{\rm eff} = -\frac{\Delta_g}{2} \sigma_z - \vec{S}\cdot\hat{\boldsymbol{\mu}}\cdot\vec B_{\rm ac}(t)~,
        \label{eq:eff_ham}
    \end{equation}
    where
    \begin{align}
        \Delta_g = 2 \gamma_s \sqrt{\sin(x)^2 (B_{x,{\rm dc}}^2+ B_{y,{\rm dc}}^2) +  \Theta(B_{z, \rm dc})^2}~,
        \label{eq:spin_state_splitting}
    \end{align}
    \begin{equation}
        \Theta(B_{z, dc}) = - B_{z, dc} \left(f\frac{\gamma_l}{\gamma_s}\cos(x) + 1\right)~,
    \end{equation}
    and the 3$\times$3 matrix $\boldsymbol{\mu}$ depends on $\vec{B}_{\rm dc} = [B_{x {\rm, dc}}, B_{y {\rm, dc}}, B_{z {\rm, dc}}]$ as well as $\xi$ (see appendix \ref{app:effective} for details). We also use $\vec{S} = \boldsymbol{\sigma}/2$, $\boldsymbol{\sigma} = (\sigma_x, \sigma_y, \sigma_z)$ ($\sigma_i$ are the Pauli matrices) and $\tan(x) = \xi / \lambda$.
    
    In the case of $\vec{B}_{\rm ac}(t) =  b(t) \vec{B}_{\rm ac} \cos(\omega t)$, where $b(t)$ changes on timescales that are slow compared to the other time-scales in the problem, $H_{\rm eff}$ in Eq.~\eqref{eq:eff_ham} can be treated in the rotating-wave approximation as (see appendix \ref{app:rwa})
    \begin{equation}
        H_{\rm eff}^{\rm RWA} = b(t) \Omega \vec{n} \cdot \boldsymbol{\sigma} 
        \label{eq:heff_rwa}
    \end{equation}
    where $\vec{n} = [(\hat{\boldsymbol{\mu}} \cdot \vec{B}_{\rm ac})_x/2 , (\hat{\boldsymbol{\mu}} \cdot \vec{B}_{\rm ac})_y/2 , 0] / \Omega$.
    If we further introduce 
    \begin{align}
        \vec{B}_{\rm dc/ac} &= B_{\rm dc/ac}
        \begin{bmatrix}
            \cos(\phi_{\rm dc/ac})\sin(\theta_{\rm dc/ac})\\
            \sin(\phi_{\rm dc/ac})\sin(\theta_{\rm dc/ac}) \\ 
            \cos(\theta_{\rm dc/ac})
        \end{bmatrix}~
    \end{align}
    we find
    \begin{align}
        &\Omega = \gamma_s B_{\rm ac} \Lambda ~,  \label{eq:RabiRate}\\
        &\begin{aligned}
            &\Lambda = \sin(x) \sqrt{
            \begin{aligned}
                &\Xi(\phi, \theta_{\rm ac})^2 \\
                &\hspace{.5cm}+ 
                \frac{\chi(\phi, \theta_{\rm dc}, \theta_{\rm ac})^2}{\sin(\theta_{\rm dc})^2 \zeta(x)^2 + \cos(\theta_{\rm dc})^2}
            \end{aligned}
            }
        \end{aligned}~,\label{eq:Lambda}
    \end{align}
    where 
    \begin{align}
        &\Xi(\phi, \theta_{\rm ac}) = \sin(\phi)\sin(\theta_{\rm ac}) \\
        &
        \begin{aligned}
        \chi(\phi, \theta_{\rm dc}, \theta_{\rm ac}) = &\sin(\theta_{\rm dc})\cos(\theta_{\rm ac})\\
        &\hspace{.5cm}- \cos(\phi)\cos(\theta_{\rm dc})\sin(\theta_{\rm ac})
        \end{aligned}
    \end{align}
    $\zeta(x) = \sin(x)/[f \cos(x) /2 + 1 ] $ and $\phi = \phi_{\rm dc} - \phi_{\rm ac}$. 
    The coupling strength $\Lambda$ provides analytical insight into the most efficient microwave control regime. 

    $H_{\rm eff}^{\rm RWA}$ can handily be interpreted as a Hamiltonian that generates rotations around the $\vec{n}$ axis on the Bloch sphere. The gate defined in Eq. \eqref{eq:gate}, then becomes in the rotating frame
    \begin{equation}
        U(t_e, t_i) = e^{\frac{i}{\hbar} \theta \vec{n} \cdot \boldsymbol{\sigma}}~,
    \end{equation}
    where $\theta = \Omega \int_{t_i}^{t_e} b(t) \d t$. Any element in $SU(2)$ (i.e. any rotation on the Bloch sphere), can be constructed by composing $U = R_x(\alpha) R_y(\beta) R_x(\gamma)$, where $R_{x/y}(\alpha)$ are rotations around the $x$ or $y$ axis on the Bloch sphere by an angle $\alpha$ \cite{hamada_minimum_2014}. Both rotational axis are accessible by either adjusting the phase of the \textit{ac} driving field $\vec{B}_{\rm ac}(t) \rightarrow f(t)\vec{B}_{\rm ac}\cos(\omega t+ \varphi)$ or by changing the orientation of $\vec{B}_{\rm ac}$ to switch between $x$ and $y$ axis (see e.g. Eq.~\eqref{eq:heff_rwa}).  We comment on numerically obtained gate fidelities of $U(t_e, t_i)$ without any approximations after discussing the control efficacy of the SiV and SnV. 
    
\section{Efficient Microwave control}

    The most efficient regime for microwave control can be determined by $\Lambda$ given in Eq.~\eqref{eq:Lambda}. 
    Having a closed expression for $\Lambda$ allows us to make some general statements about the efficiency of control for different strain regimes and orientations of $\vec{B}_{\rm ac/dc}$. We first analyze general properties of $\Lambda$ as a function of $x$ and then discuss $\Lambda$'s dependence on the orientations of $\vec{B}_{\rm dc}$ and $\vec{B}_{\rm ac}$. 
    
    Notably, Eq.~\eqref{eq:Lambda} only depends on the difference of the azimuthal angles $\phi_{\rm dc} - \phi_{\rm ac}$, the polar angles $\theta_{\rm dc}$, $\theta_{\rm ac}$ and the dimensionless variable $x \in \{0,\pi/2\}$. The fact that $\xi$ solely enters through its comparison with the spin orbit interaction $\lambda$ is another interesting feature of Eq.~\eqref{eq:Lambda}, which allows for a general discussion of the interplay of strain, the Jan-Teller effect and the spin orbit interaction, completely independent of the G4V.   
    
    With very little effort we can assert that $\Lambda = 0$ for $\vec{B}_{\rm ac}$ parallel $\vec{B}_{\rm dc}$, which is in line with physical intuition: no driving of the spin-qubit is possible in the absence of spin-mixing.   
    Before investigating the local extrema of $\Lambda$, we analyze the limits $\xi \rightarrow 0$ and $\xi \rightarrow \infty$, which correspond to $x = 0$ and $x = \pi/2$, respectively. 
    
    For small values $x \rightarrow 0$, we expand $\Lambda$ in leading orders of $x$: 
    \begin{equation}
        \lim_{x \rightarrow 0} \Lambda = f(\phi, \theta_{\rm ac}, \theta_{\rm dc}) x + \mathcal{O}(x)^2
    \end{equation}
    where $f(\phi, \theta_{\rm ac}, \theta_{\rm dc})$ can be found in appendix \ref{app:efficient}.
        
    It would appear that $\Lambda = 0$ for $x = 0$, corresponding to an absence of the Jahn-Teller interaction, as well as internal and external strain. Interestingly, as we will explain in the next paragraphs, this is not true for all directions of the magnetic field so that $\Lambda \neq 0$ for $x = 0$ under certain conditions. 

    For $x \rightarrow \infty$, we find that 
    \begin{equation}
        \lim_{\xi \rightarrow \infty}\Lambda =  g(\phi, \theta_{\rm dc}, \theta_{\rm ac})            
    \end{equation}
    where $g(\phi, \theta_{\rm dc}, \theta_{\rm ac})$ only depends on the magnetic fields' orientations (see appendix \ref{app:efficient}), sometimes referred to as the free electron limit \cite{rosenthal_microwave_2023}.
    Notably, this is the case for either $\xi \rightarrow \infty$ or $\lambda \rightarrow 0$, which means that 
    as long as the strain outcompetes the spin-orbit coupling, the efficiency of the driving will purely depend on the relative orientation of $\vec{B}_{\rm dc}$ and $\vec{B}_{\rm ac}$.     
        
    Local extrema of $\Lambda$ as a function of $\phi$, $\theta_{\rm dc}$ and $\theta_{\rm ac}$ can be found analytically. The first extremum of interest is $\phi = 0$ deg$,~\theta_{\rm dc} = 0$ deg and $\theta_{\rm ac} = 90$ deg, 
    which corresponds to $\vec{B}_{\rm dc}$ being parallel to the main symmetry axis of the G4V, and $\vec{B}_{\rm ac}$ being oriented in the $xy$-plane. For this combination of angles $\Lambda_1 = \sin(x)$.
    
\begin{figure*}[ht!]
        \centering
       \includegraphics[width=.8\textwidth]{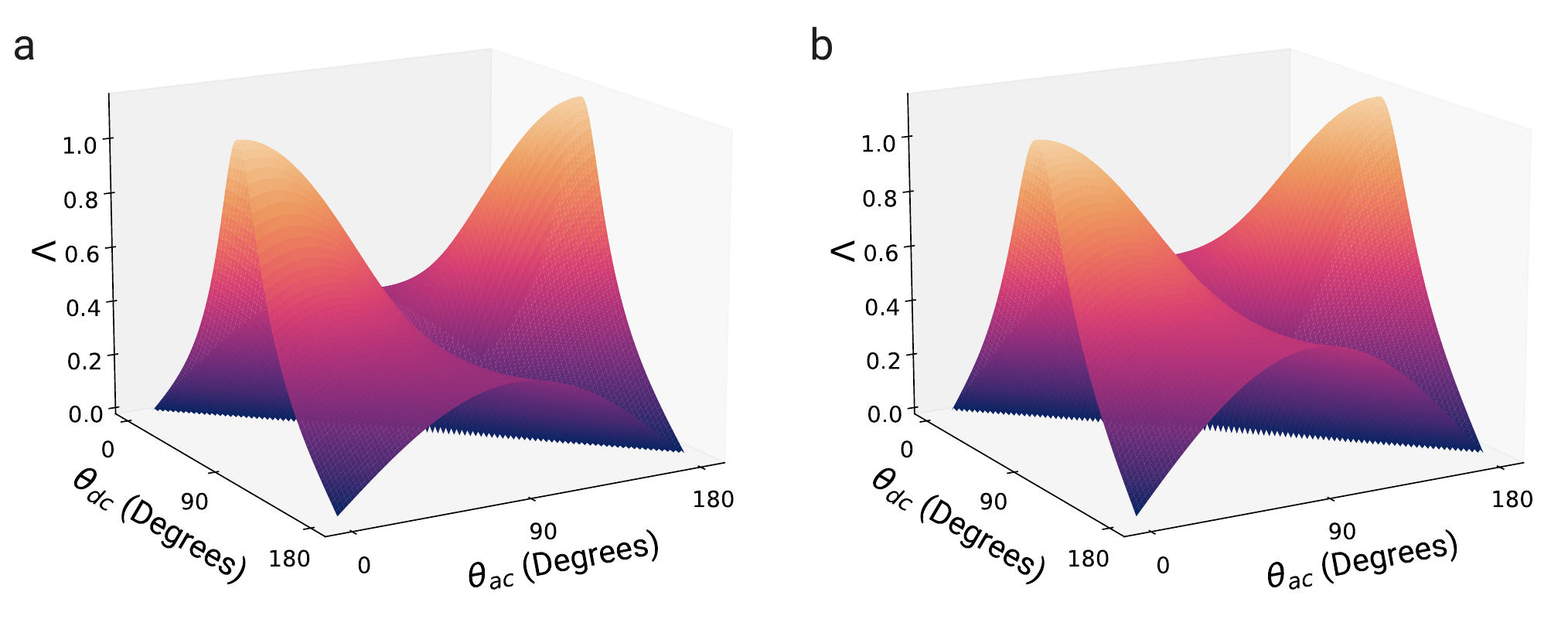}
        \caption{$\Lambda$ as a function of the magnetic field angles $\theta_{\rm dc}$ and $\theta_{\rm ac}$ unstrained (a) and strained ($E_x=0.2\,\times10^{-3}$) (b) for the SnV, $B_{\rm dc}= 100$ mT.}
        \label{fig:Rabi function of DC_angle and AC_angle zero strain}
    \end{figure*}

    For infinite strain, i.e. $x = \pi/2$, the coupling is maximized $\Lambda_1 = 1$ and $\Lambda_1 = 0$ for zero strain and zero Jahn Teller interaction, which is in accordance with previous reports on microwave control \cite{rosenthal_microwave_2023, guo_microwave-based_2023}.   
    In the case of $\xi < \lambda$ we can reproduce the finding in \cite{maity_coherent_2020}, where $\Omega = \gamma_s B_{\rm ac}  \sqrt{\upsilon_x^2 + \upsilon_y^2} /\lambda$.    
    One can show that this extremum is a saddle point as a function the two polar angles of the magnetic fields (see Fig. \ref{fig:Rabi function of DC_angle and AC_angle zero strain}), which means there exist orientations, that allow for more efficient control. The difference of the azimuths $\phi$ has little impact on the steepness of the saddle point, which means that changing the azimuthal orientation of the fields will have little impact on the driving efficiency.  
    
    The second notable local extremum is given by $\phi = 0$ deg$, \theta_{\rm dc} = 90$ deg$, \theta_{\rm ac} = 0$ deg, which corresponds to $\vec{B}_{\rm dc}$ lying in the $xy$-plane and $\vec{B}_{\rm ac}$ being oriented in parallel to the symmetry axis of the system. For this field configuration
    \begin{align}
        \begin{aligned}
            \Lambda_2 & = \sin(x) |1/\zeta(x)| \\
                    & = f\cos(x)/2 + 1~,
        \end{aligned}
        \label{eq:no_strain_max}
    \end{align}
    which is maximized for $x = 0$, meaning for either $\xi \ll \lambda$ or no contribution from either the Jan-Teller interaction or internal and external strain. At $x = 0$, $\Delta_g = 0$, which means that the spin-qubit is energetically degenerate, which would make the spin-qubit optically inaccessible. This, however, is not physical, since there will always be a Jahn-Teller contribution so that $\Delta_g \neq 0$. Additionally, for $\Lambda_2$, $\Delta_g = 2\gamma_s |\sin(x)| B_{\rm dc}$, the splitting can be increased by increasing $B_{\rm dc}$.      
    
    Contrary to $\Lambda_1$, $\Lambda_2$ is a global extremum as a function of both $\phi$ and the polar angles of the fields. This means that for the most efficient microwave control $\Lambda_2$ is the optimum. Their relative magnitude is given by $\Lambda_2/ \Lambda_1 = f \cot(x)/2 + 1/\sin(x)$. Depending on $x$ the magnetic field configuration responsible for $\Lambda_2$ can be arbitrarily more effective for controlling the spin-qubit. 
    
    We observe that to leading order $\lim_{x\rightarrow0}(\Lambda - \Lambda_2)  \propto - (\lambda^4/\xi^4) \theta_{\rm dc}^2$ for $\theta_{\rm ac} = {\rm constant}$, which means that the slope close to $\Lambda_2$ becomes infinitely steep for vanishing $\xi$. For emitters with $\lambda < \xi$, the local extrememum is therefore less pronounced. Because $\lambda_{\rm SnV}, \lambda_{\rm PbV} \gg  \lambda_{\rm SiV }$ more strain should lead to a more forgiving parameter regime for the magnetic fields for efficient spin control of the SnV and PbV. The narrow extremum for larger values of $\lambda$ is also the most likely explanation why $\Lambda_2$ so far has evaded experimental detection. 
    The steepness is bound by the Jahn-Teller contribution to $\xi$ and therefore $\xi \neq 0$ for any of the G4V. For the SnV $\Omega$ is shown in Fig.~\ref{fig:Rabi function of DC_angle and AC_angle zero strain}. We discuss an optimal control regime in the next section. 

    Note that $\Lambda > 1$ due to the contribution of the orbital Zeeman effect when $\vec{B}_{\rm dc}$ is aligned perpendicular to the symmetry axis and $\vec{B}_{\rm ac}$ is aligned parallel to it. We show the magnitude of the amplification in appendix~\ref{app:Amplification}.

\section{Optimal control regime}

    The question of which orientation of the magnetic fields are optimal is highly dependent on the system's intended use in an application. For most quantum applications, such as the use of the electronic spin as a memory \cite{nguyen_integrated_2019}, as a nuclear spin interface \cite{metsch_initialization_2019} or as a mediator of entanglement between photons \cite{lee_quantum_2019} the electronic spin has to be optically addressable, initialized, controlled and read out with high fidelity. Another important aspect is the spin-qubit's coherence time, which also depends on the \textit{dc} magnetic field orientation \cite{rogers_all-optical_2014}.   
    If the \textit{dc} magnetic field axis is fixed for a given application, then all of these factors have to be considered. In this section we provide a brief discussion of the important quantities that impact the optimal choice of orientation of the magnetic field.

    Optical addressability depends on the difference of $\Delta =| \Delta_e - \Delta_g|$, where $\Delta_g$ and $\Delta_e$ are shown in Fig.~\ref{fig:fig0}. They can be calculated by replacing $\lambda \rightarrow \Lambda_{e,g}$ in \eqref{eq:spin_state_splitting}. We adopt the convention in \cite{nguyen_integrated_2019} that optically addressability requires $\Delta > \Gamma$, where $\Gamma$ is the faster radiative decay rate of the two transitions. This rate depends, of course, on external parameters such as the local density of electromagnetic modes and can be highly influenced by the presence of, for example, a surface boundary or a cavity \cite{li_coherent_2015}.    
    
    Here we assume lifetime-limited linewidths, which is reasonable for the SnV at $4$~K and the SiV at $200$~mK. In the presence of homogeneous broadening due higher temperatures \cite{jahnke_electronphonon_2015} or inhomogenous broadening 
    due to fluctuations in the charge environment \cite{orphal2023, pieplow_quantum_2024} the above condition has to be modified: $\Delta > \gamma_{\rm lw}$, where $\gamma_lw$ now is the broadened linewidth of the respective transition.
    
    The splitting is maximized for the $\theta_{\rm dc} = 90$ deg orientation and non-trivially depends on the strain, see Fig.~\ref{fig:fig0c}. As long as the splitting does not vanish, it can always be increased by increasing the strength of the magnetic field.   

    The ability to initialize the system is highly dependent on the details of the initialization scheme. Here we briefly discuss two approaches to the spin initialization: Spin pumping \cite{rogers_all-optical_2014} and cavity scattering \cite{nguyen_integrated_2019}.  
    \begin{figure}
        \centering
        \includegraphics[width = .8\columnwidth]{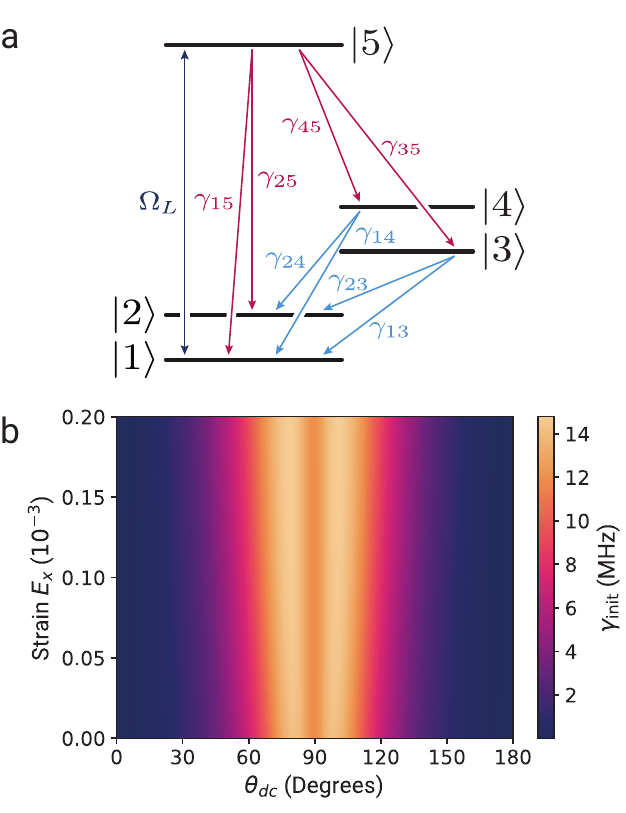}
        \caption{Spin state initialization rate of  the SnV defect. a) Level scheme showing the optical (red) and phononic (light blue) relaxation processes $\gamma_{ij}$. The transition $1\leftrightarrow 5$ is resonantly driven by a laser $\Omega_L$. b) Initialization rate $\gamma_{\rm intit.}$ as a function of Strain. For most $E_x$ values, the highest initialization rates are achieved for $\vec B_{\rm dc}$ ($B_{\rm dc}=100$ mT) oriented slightly away from the plane orthogonal to the symmetry axis.}
        \label{fig:initialization}
    \end{figure}
    Spin pumping requires the spin initialization rate $\gamma_{\rm init} \gg 1/T_{1}$, where $T_1$ is the spin relaxation time of the qubit \cite{rogers_all-optical_2014}. For resonant spin pumping we can calculate $\gamma_{\rm init}$ as a function of the relevant spontaneous emission rates and phononic decay processes, as shown in Fig.~\ref{fig:initialization}. 
    We use a rate model in the regime of strong pumping $\Omega_L \gg \gamma_{15} + \gamma_{25} + \gamma_{35} + \gamma_{45}$, where the rates $\gamma_{ij}$ are illustrated in Fig. \ref{fig:initialization} and $\Omega_L = |\langle 1 |\vec{p}\cdot\vec{E}|5\rangle |/ 2\hbar$ and $\vec{E}$ is the electric field strength of a laser driving the system at the frequency that is resonant with the transition energy $\hbar \omega_{3\downarrow,1\downarrow}$.$~\Omega_L$ encodes the cyclicity, which is important for the spin readout. From the rate equations corresponding to Fig.~\ref{fig:initialization}a (see appendices \ref{app:radiative emission} and \ref{app:rate_equations}) we find the quasi stationary solution in the strong pumping limit
    \begin{align}
        \rho_{11} &= e^{-\gamma_{\rm init}t} \label{eq:pop11}/2\\
        \rho_{22} &= 1 - e^{-\gamma_{\rm init}t} \label{eq:pop22}/2
    \end{align}
    where 
    \begin{equation}
    \label{eq:gamma init}
    \gamma_{\rm init} = \frac{\gamma_{25}}{2}\left(1 + \frac{\gamma_{35}/\gamma_{25}}{1 + \gamma_{13}/\gamma_{23}} + \frac{\gamma_{45}/\gamma_{25}}{1 + \gamma_{14}/\gamma_{24}} \right)
    \end{equation}
    and we assume that the initial state is the mixed state $\rho = \frac{1}{2}(|1\rangle\langle 1| + |2\rangle\langle 2|)$. The expression for $\gamma_{\rm init}$ shows that if the spin non-conserving transitions vanish initialization is no longer possible, which is true when $\vec{B}_{\rm dc}$ is parallel to the defect's symmetry axis. For spin pumping to be efficient, an off-axis field is required, so that $\gamma_{25}, \gamma_{45}, \gamma_{23}, \gamma_{24} \neq 0$. For $\vec{B}_{\rm dc}$ orthogonal to the symmetry axis, this is always the case. We show the angular and strain dependence in Fig.~\ref{fig:initialization}b. 

    Spin initialization using a cavity scattering scheme was shown to be successful in \cite{nguyen_integrated_2019}, for both parallel and orthogonal orientation of $\vec{B}_{\rm dc}$. If it comes to initialization there is thus no fundamental reason to not choose the orthogonal orientation of $\vec{B}_{\rm dc}$, when it comes to spin initialization.  
    
    Resonant single-shot read out schemes \cite{sukachev_silicon-vacancy_2017} work best, when there is a spin conserving transition, that can be cycled continuously, without inducing a spin flip. The finite branching ratio $\gamma_{52}/\gamma_{51}$ for the orthogonal configuration of $\vec{B}_{\rm dc}$, is therefore not ideal for a resonant single-shot readout, and then becomes highly dependent on, for example, detection efficiencies of an optical setup. The cavity scattering scheme, does not have these issues, but requires a cavity-color center system in a specific coupling regime to function \cite{bhaskar_experimental_2020}.    

    Decoherence due to electron-phonon interactions may present a challenge for the most efficient magnetic field orientations, because the qubit state levels can be directly coupled through these interactions. A dependence of the spin relaxation time on the \textit{dc} magnetic field orientation has been observed in \cite{rogers_all-optical_2014}, dramatically decreasing $T_2 = 2.4$ ms to  $T_2 = 3.4$ $\mu$s. However, these measurements have been conducted at high temperature $> 4$ K were the phonon's thermal occupations is significantly increased at the relevant energies. To address this concern, we calculated (see appendix \ref{app:Phonon}) that within an orbital branch, for example $|1\rangle$ and $|2\rangle$), the electron-phonon coupling is comparatively small and restricted to the $\epsilon_{xy}$ phononic mode. At low temperatures or an adequate design of the surrounding nanostructure featuring a phononic band gap, the $\epsilon_{\rm xy}$ can be minimized. On the other hand, an increased coupling to phononinc modes can also be used to engineer an efficient mechanical interface \cite{maity_coherent_2020, clark_nanoelectromechanical_2023}.

\section{Microwave control of the SiV and SnV}
    
    In this section we present some concrete values for the obtainable Rabi frequencies for the SiV and the SnV, two vacancies for which microwave driving has been shown to work experimentally. For both vacancies' parameters we use Table \ref{table:SiV SnV parameters}.
    \begin{figure*}[ht!]
        \centering
        \includegraphics[width=.8\textwidth]{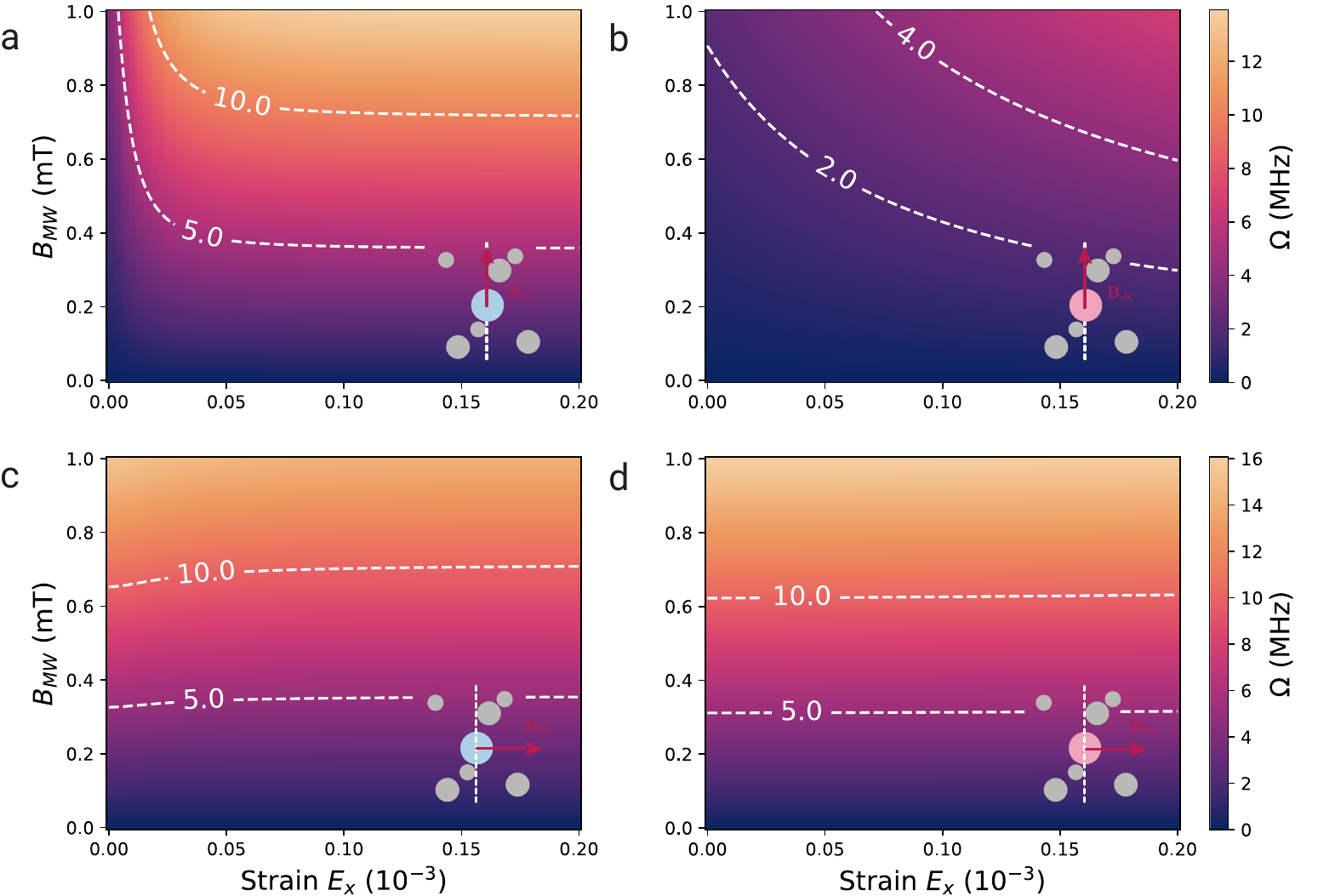}
        \caption{Rabi frequency $\Omega$ dependence on Strain component $E_x$ (and $\epsilon_{xy} =0$) and  microwave field strength for SiV (a, c) and SnV (b, d). (a, b) depict the configuration where $B_{\rm dc}$ is parallel and $B_{\rm ac}$ orthogonal to the symmetry axis, while (c, d) depict the configuration where $B_{\rm dc}$ is perpendicular and $B_{\rm ac}$ is parallel to the symmetry axis) with $B_{\rm dc}=100$ mT.}
        \label{fig:Rabi_Frequency B Strain}
    \end{figure*}
    \begin{figure*}[ht!]
        \centering
        \includegraphics[width=.8\textwidth]{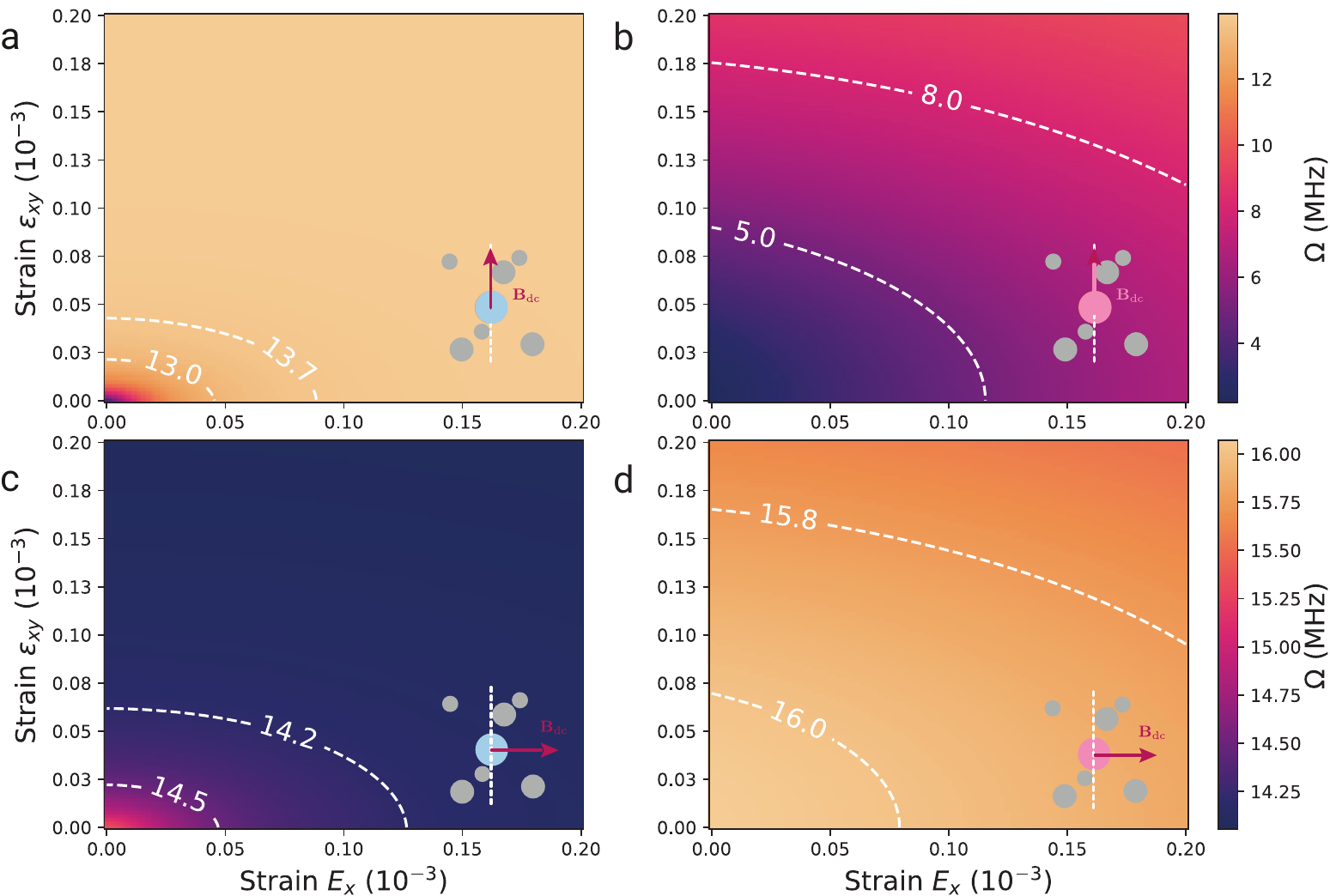}
        \caption{Rabi Frequency $\Omega$ dependence on Strain contributions $E_x$ and $\epsilon_{xy}$ for SiV (a, c) and SnV (b, d). (a, b) depict the configuration where $B_{\rm dc}$ is parallel and $B_{\rm ac}$ orthogonal to the symmetry axis, while (c, d) depict the configuration where $B_{\rm dc}$ is perpendicular and $B_{\rm ac}$ is parallel to the symmetry axis) with $B_{\rm dc}=100$ mT. The highest color axis values for the parallel configuration (a, b) is $\Omega=28$ MHz, which is the lowest value for the perpendicular configuration (c, d).}
        \label{fig:Rabi_Frequency Strain Strain}
    \end{figure*}
    For the case of $\vec{B}_{\rm dc}$ parallel to the symmetry axis, Fig.\,\ref{fig:Rabi_Frequency B Strain} shows the dependence of the Rabi frequency $\Omega$ as a function the microwave field strength and the externally applied strain for the SiV and SnV.
    The strain direction $\epsilon_{xy}$ shows a stronger effect in Fig. \ref{fig:Rabi_Frequency Strain Strain} due to the factor $2$ in $H_{\rm strain}$ [Eq.~\eqref{eq:Strain}]. 

    Remarkably, for the chosen values of $B_{\rm ac} = 1$ mT and $B_{\rm dc} = 100$ mT as well as zero strain, the SnV exhibits an increased $\Omega_{\rm SnV} = 16.1$ MHz compared to the SiV $\Omega_{\rm SiV} = 15.3$ MHz, when $\vec{B}_{\rm dc}$ is perpendicular and $\vec{B}_{\rm ac}$ parallel to the symmetry axis. This configuration allows for more efficient control of the SnV compared to the SiV, which is notable given the contrasting behaviour observed when $\vec{B}_{\rm dc}$ aligns with the symmetry axis.    
    
    As has been noted in previous work \cite{rosenthal_microwave_2023, guo_microwave-based_2023}, increasing the Rabi frequency requires increased strain. As predicted by Eq.~\eqref{eq:Lambda}, the Rabi frequency saturates as a function of strain at around $\Omega_{\rm max} = 14$ MHz for $B_{\rm ac} = 1$ mT and $B_{\rm dc} = 100$ mT. 
    Due to the reduced $\lambda$ for the SiV, we observe the same behavior as for the SnV except that Rabi frequency increases faster with more strain and more quickly saturates, as shown in Fig.~\ref{fig:Rabi_Frequency B Strain}. 

\begin{figure}[ht!]
        \centering
        \includegraphics[width=0.9\columnwidth]{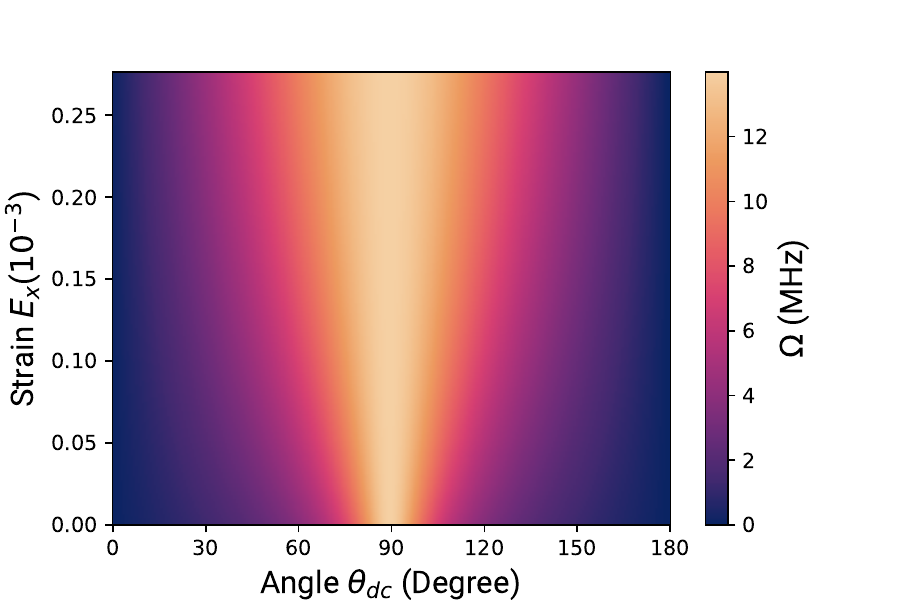}
        \caption{Rabi frequency as a function of $\theta_{\rm dc}$ and Strain component $E_x$ ( and $\epsilon_{xy} =0$) with $B_{\rm dc}=100$ mT.}
        \label{fig:Rabi as a function of DC angle and Strain for AC angle at zero}
    \end{figure}

    Fig. \ref{fig:Rabi as a function of DC angle and Strain for AC angle at zero} shows the dependence of $\Omega$ on the polar angle of the magnetic field ($\vec B_{\rm ac}$ is parallel to the symmetry axis). For larger values of strain the system is less sensitive to misalignment of $\theta_{\rm dc}$. 

    Finally, we numerically calculate microwave qubit gate fidelities for the SiV and SnV. We use an averaged gate fidelity (see for example \cite{magesan_gate_2011}), which we define as
    \begin{equation}
        F=\frac{1}{4\pi} \int_{\mathcal{S}} {\rm Tr}(V|\psi\rangle\langle \psi |V^\dagger U|\psi\rangle\langle \psi |U^\dagger) \d\psi
        \label{eq:fid}
    \end{equation}
    where $V$ is the target gate and $U$ the numerically exact gate, calculated by numerically integrating Schr\"odingers equation. Here we consider rotations around the x-axis.

    In \eqref{eq:fid}, $|\psi\rangle = \cos(\frac{\theta}{2}) |1\rangle + e^{\i \phi} \sin(\frac{\theta}{2})|2\rangle$, for which $\theta \in [0,\pi]$ and $\phi \in [0,2\pi]$. The integral is calculated by integrating over the surface of the Bloch sphere $\mathcal{S}$, using the surface element $\d\psi = \sin(\theta) \d\theta\d\phi$. Gate durations are  $\frac{1}{\Omega}$ for a $\pi$ rotation and $\frac{1}{2\Omega}$ for $\pi/2$ rotation. 
    The infidelities $1-F$ for SiV and SnV rotation gates are given in Table \ref{table:fidelity}. 

    Orbital splitting in group-IV vacancies is mainly due to the spin-orbital interaction and the additional strain contribution. The approximations leading to Eq.~\eqref{eq:RabiRate} requires that $\Delta_{\rm g}$ is much smaller than the orbital splitting. This condition does not hold for SiV in the case of zero strain as SiV spin-orbital coupling is only one order of magnitude bigger than $\Delta_{\rm g}$. Using the gate duration derived from Eq.~\eqref{eq:RabiRate} therefore leads to slightly decreased gate fidelities. We choose to optimize $B_{ac}$ and therefore the gate duration to minimize the $\pi$ and $\pi/2$ rotation infidelities for the SiV shown in Table \ref{table:fidelity}. We checked, that the addition of strain to the SiV, which increases the splitting of the orbital branches, drastically improved the fidelities without any optimization. Such an optimization was also unnecessary for the SnV, due to the already much larger spin-orbit interaction. In conclusion, Table~\ref{table:fidelity} demonstrates that achievable fidelities are limited only by operational errors and experimental imperfections.
    \begin{table}[ht!]
    \centering
    \begin{tabular}{c c c c c c} 
      & $B_{\rm ac}$ (mT) & $\Omega$ (MHz) &  $T_{\rm \pi/2}$ (ns) & $T_{\rm \pi}$ (ns) &  $1-F$ \\
    \hline
    \hline
    \multicolumn{5}{c}{SiV} \\
    \hline
     $B_{dc_\parallel}$ & 3.7 & 12.9 & 19.3 & 38.6 & $< 10^{-5}$ \\
     $B_{dc_\perp}$ & 3.7 & 56.4 & 4.4 & 8.9 &  $< 10^{-4}$ \\
     \hline
    \multicolumn{5}{c}{SnV} \\
    \hline
      $B_{dc_\parallel}$ & 1.0 & 2.2 & 113.4 & 226.8 & $< 10^{-7}$ \\
     $B_{dc_\perp}$ & 1.0 & 16.1 & 15.6 & 31.1 &  $< 10^{-8}$ \\[1ex] 
     \hline
     \hline
    \end{tabular}
    \caption{Infidely $1-F$ of $\pi/2$ and $\pi$ gate for the SiV and SnV at zero strain and without considering decohering processes and without experimental control imperfections. The values are shown for both configuration ($\vec{B}_{dc_\parallel}$ and $\vec{B}_{dc_\perp}$) for $B_{dc} =200~\rm mT$. $B_{\rm ac}$ is the \textit{ac} magnetic field strength, $\Omega$ is the Rabi frequency, $T_{\rm \pi/2}$ and $T_{\rm \pi}$ are the duration of $\pi/2$ and $\pi$ rotation around the x-axis on the Bloch sphere, respectively. For both the SiV and SnV $\Delta_g \gg 1/\tau_{\rm rad}$ at $B_{\rm dc} > 0.1$ T, where $\tau_{\rm rad} > 4.5 (1.7)$ ns is the radiative lifetime of the SnV 
    \cite{trusheim_transform-limited_2020}(SiV \cite{tamarat_stark_2006}).}
    \label{table:fidelity}
    \end{table}

\section{Conclusion \& Discussion}

    In this work we theoretically identified and analysed the most efficient regime for microwave control in the presence of different strain regimes for G4V color centers in diamond. We demonstrated that independent of the specific system parameters of the type of G4V, orienting $\vec B_{\rm dc}$ perpendicular to the symmetry axis of the vacancy and driving the system with $\vec B_{\rm ac}$ oriented in parallel is the most efficient configuration independent of the magnitude of the strain. This is extremely advantageous for controlling the heavier element color centers such as the SnV and the PbV, which suffer from a bigger penalty in terms of efficiency for a non-perpendicular field orientation of $\vec B_{\rm dc}$, requiring higher microwave power, which can potentially increase heating of a sample and have other technical disadvantages.
    Our analysis also implies that, as far as microwave control is concerned, strain tuning is not necessarily required, greatly reducing experimental overheads. 

    In conclusion we presented a novel and highly efficient path for controlling G4V independent of strain and the spin-orbit coupling that is compatible with all the required quantum information processing builing blocks, such as initialization, coherent control, and readout. 

\section*{Acknowledgements}

The authors would like to thank David Hunger, Eric I. Rosenthal, Yannick Strocka, Fenglei Gu and Thiering Gergő for their scientific input. Moreover, funding for this project was provided by the European
Research Council (ERC, Starting Grant project QUREP, No. 851810) and the German Federal Ministry of Education
and Research (BMBF, project QPIS, No. 16KISQ032K; project QPIC-1, No. 13N15858).

\section*{Author Contributions}

M.B compiled the model and performed the comprehensive numerical analysis of the interplay between strain and magnetic field orientations, gate fidelities, initialization and readout. G.P. performed the analytical analysis of the microwave driving as well as the initialization and readout.     
T.S. and G.P developed the idea and supervised the project. All authors contributed to the writing of the manuscript. 

\begin{appendix}

\section{Derivation of effective Hamiltonian}
We perform the adiabatic elimination using the formalism that can be found in \cite{paulisch_beyond_2013}. The Hamiltonian $H(t) = H_{\rm dc} + H_{\rm ac}(t)$ is grouped according to:
 \begin{align}
     H = 
     \begin{pmatrix}
     \boldsymbol{\omega} & \boldsymbol{\Omega}/2 \\
     \mathbf{\Omega^\dagger}/2 & \mathbf{\Delta}
     \end{pmatrix} ~,
 \end{align}
where $\boldsymbol{\omega}$ operates on $|-, \downarrow\uparrow\rangle$, $\mathbf{\Delta}$ operates on $|+, \downarrow\uparrow\rangle$ and $\mathbf{\Omega}$ connects the two sectors. We do not detail the components here. They are straightforward to calculate using the description in the main text. Formally, the elimination is performed according to 
\begin{equation}
    H_{\rm eff} = \boldsymbol{\omega} - \boldsymbol{\Omega}\frac{1}{4\boldsymbol{\Delta}}\boldsymbol{\Omega}^\dagger~.
\end{equation}
We furthermore neglect $\boldsymbol{\Omega}(1/4\boldsymbol{\Delta}) \boldsymbol{\Omega}^\dagger$ leaving us with $H_{\rm eff} = \boldsymbol{\omega}$, which is appropriate as long as $\boldsymbol{\Omega}$ only weakly couples the states $|-, \downarrow\uparrow\rangle$ and $|+, \downarrow\uparrow\rangle$ and is small compared to $\boldsymbol{\Delta}$. The latter assumption works better for the SnV, due to the larger $\lambda$ (see Table~\ref{table:SiV SnV parameters}).

After the adiabatic elimination the effective Hamiltonian becomes
 
 \label{app:effective}
    \begin{equation}
        H_{\rm eff} = H_{\rm B_{\rm dc}}^{\rm eff} + H^{\rm eff}_{\rm B_{\rm ac}(t)}
    \end{equation}
    where 
    \begin{align}
        &
        \begin{aligned}
            H_{\rm B}^{\rm eff} = & \gamma_s[\Phi(B_x, B_y) \sigma_x + \Phi(-B_y, B_x) \sigma_y+ \Theta(B_z) \sigma_z] 
        \end{aligned}
        \\
        &
        \begin{aligned}
            &\Phi(B_x, B_y) = \frac{1}{\xi^2}\left(B_x \upsilon_y^2 + B_y \upsilon_y \lambda \vphantom{\frac{A}{B}} 
            \right.
            \\
            &\hspace{2.5cm}\left.+\frac{\lambda\upsilon_x}{\sqrt{\lambda^2 + \xi^2}}(- \upsilon_y B_y + \lambda B_x)\right)~,
        \end{aligned}\\
        & \Theta(B_z) = - B_z \left(\frac{\lambda f \gamma_l}{\gamma_s\sqrt{\lambda^2+ \xi^2}} + 1\right)~ 
    \end{align}
    where $\sigma_i$ are the Pauli matrices. 
    To understand the interplay of strain and microwave control we express the above Hamiltonian in terms of 
    the energy eigenstates of $H_{\rm B_{\rm dc}}^{\rm eff}$.
    Finally, we can express $H_{\rm B_{\rm ac}(t)}$ in terms of the new eigenbasis to calculate the effective Hamiltonian in Eq.~\eqref{eq:eff_ham} for which we introduce $\vec{B}_{\rm \perp, dc} = (B_{x, \rm dc}, B_{y, \rm dc})$.
    The interaction with the time dependent magnetic field in Eq.~\eqref{eq:eff_ham} is then mediated by
    \begin{equation}
        \hat{\boldsymbol{\mu}}=
        \begin{pmatrix}
            \mu_{xx} & \mu_{xy} & \mu_{xz} \\
            \mu_{yx} & \mu_{yy} & \mu_{yz} \\
            \mu_{zx} & \mu_{zy} & \mu_{zz} 
        \end{pmatrix}~,
    \end{equation}
    \begin{align}
        \mu_{xx} & = 4 \gamma_s^2 \sin(x) \frac{B_{x, \rm dc}\Theta(B_{z, \rm dc})}{B_{\perp, \rm dc}\Delta_g}~,  \\
        \mu_{yx} & = 2\gamma_s \sin(x) \frac{B_{y, \rm dc}}{B_{\perp, \rm dc}}~,\\
        \mu_{zx} & = 4\gamma_s \sin(x)^2\frac{B_{x, \rm dc}}{\Delta_g}~,\\
        \mu_{xy} & = \frac{B_{y, \rm dc}}{B_{x, \rm dc}} \mu_{xx}~,\\
        \mu_{yy} & = -\frac{B_{x, \rm dc}}{B_{y, \rm dc}} \mu_{yx}~,\\
        \mu_{zy} & = \frac{B_{y, \rm dc}}{B_{x, \rm dc}} \mu_{zx}~, \\
        \mu_{xz} & = - 4\gamma_s \sin(x) \frac{ B_{\perp, \rm dc}\Theta(B_{z, \rm dc})}{B_{z, \rm dc} \Delta_g}~, \\
        \mu_{yz} & = 0 ~,\\
        \mu_{zz} & = 4\gamma_s^2\frac{\Theta(B_{z, \rm dc})^2}{B_{z, \rm dc} \Delta_g} 
    \end{align}

\section{Amplification effect}
\label{app:Amplification}

\begin{figure}[ht!]
    \includegraphics[width=.9\columnwidth]{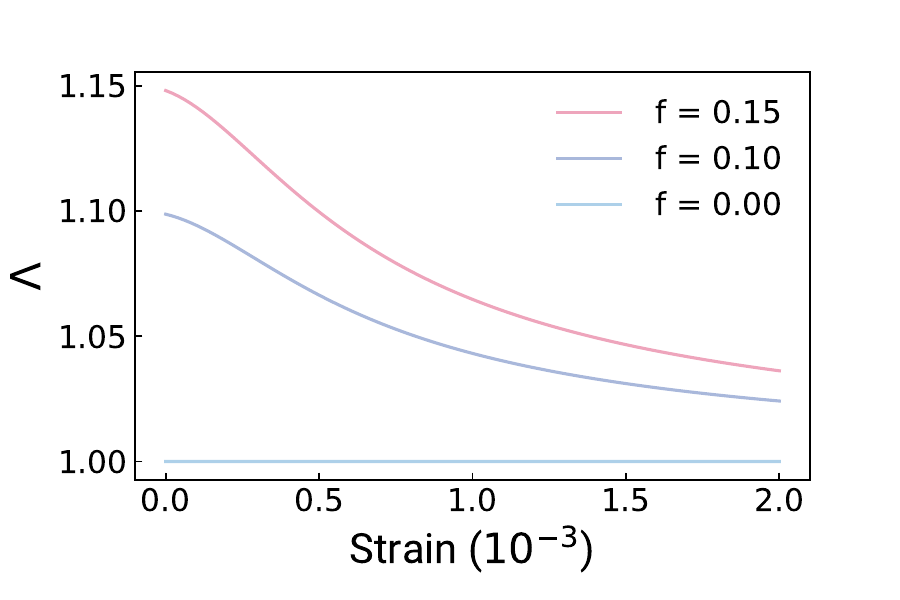}
    \caption{The strain dependence of $\Lambda$ is shown for $\vec{B}_{\rm ac}$ oriented parallel to the symmetry axis and $\vec{B}_{\rm dc}$ orthogonal to the symmetry axis for three values of $f$ to demonstrate the amplification effect due to the orbital Zeeman interaction.}
    \label{fig:Amplification}
    \end{figure}

In the perpendicular magnetic field configuration $\Lambda>1$, which means there is an amplification effect. This effect, as shown in Fig.~\ref{fig:Amplification}, arises from the orbital Zeeman interaction.

\section{Rotating Wave Approximation}
\label{app:rwa}

 \begin{equation}
        H_{\rm eff} = -\frac{\Delta_g}{2} \sigma_z - \vec{S}\cdot\hat{\boldsymbol{\mu}}\cdot\vec B_{\rm ac}(t)~.
        \label{eq:eff_hamApp1}
    \end{equation}

The first term in equation \ref{eq:eff_hamApp1} is the free evolution of the system. We can eliminate this term in a rotating frame defined by the transformation

 \begin{align}
         |\Tilde{\Psi}\rangle &= U |\Psi\rangle,\\
        U &= e^{-i\frac{\Delta_g}{2} t \sigma_z}.
        \label{eq:Rotating frame}
 \end{align}

The new Hamiltonian in this rotating frame is then given by
 \begin{equation}
        \Tilde{H} = U H_{\rm eff} U^{\dag}-iU \dot{U}^{\dag}
        \label{eq:eff_hamApp}
    \end{equation}
In the rotating frame the Hamiltonian becomes
 \begin{equation}
        \Tilde{H} = b(t)
        \begin{pmatrix}
            0 & \tilde{\Omega}\left(1+e^{-2i\Delta_g t}\right) \\
            \tilde{\Omega}^*\left(1+e^{2i\Delta_g t}\right) & 0
        \end{pmatrix}~,
        \label{eq:app_rwaderiv}
\end{equation}
where rotating wave approximation states entails neglecting the terms $e^{\pm 2i\Delta_gt}$, which is justified as long as $\Delta_g$ is greater than any other energy scale in the system. In Eq.~\eqref{eq:app_rwaderiv} the complex $\tilde{\Omega}$ can be inferred from Eq.~\eqref{eq:heff_rwa}. 

\section{Efficient control}
\label{app:efficient}

The two functions $f(\phi, \theta_{\rm ac}, \theta_{\rm dc})$ and $g(\phi, \theta_{\rm dc}, \theta_{\rm ac})$  can be found by taking the respective limits $\lim{\Lambda}_{x\rightarrow 0, \pi/2}$. 
We find 
\begin{align}
    f(\phi, \theta_{\rm dc}, \theta_{\rm ac}) &= \sqrt{
    \begin{aligned}
    & \sin(\phi)^2\sin(\theta_{\rm ac})^2 +\\ 
    & [\cos(\phi)\sin(\theta_{\rm ac})-\cos(\theta_{\rm ac})\tan(\theta_{\rm dc})]^2
    \end{aligned}
    }
\\
     g(\phi, \theta_{\rm dc}, \theta_{\rm ac}) &=\sqrt{\Xi(\phi, \theta_{\rm ac})^2+ \chi(\phi, \theta_{\rm dc}, \theta_{\rm ac})^2}
\end{align}

\section{Electron-phonon interaction}
\label{app:Phonon}

The thermal phonon bath in diamond interacts with the color center and causes decay and decoherence. This interaction can be modeled by interpreting phonons as dynamical strains deformations \cite{lemonde_phonon_2018}
\begin{equation}
    H_{\rm e-p}= d E_x L_{E_x} + 2d \epsilon_{xy} L_{\epsilon_{xy}}~,
\end{equation}
where $E_x$ and $\epsilon_{xy}$ are now dynamical fields. $L_{E_x}=|e_{gx}\rangle \langle e_{gx}|-|e_{gy}\rangle \langle e_{gy}|$ and $L_{\epsilon_{xy}}=|e_{gx}\rangle \langle e_{gy}|+|e_{gy}\rangle \langle e_{gx}|$. The operators $L_{E_x}$ and $L_{\epsilon_{xy}}$ determine the levels coupled by electron-phonon interaction. \\

Decay rates due to electron-phonon interaction can be computed using Fermi's golden rule
\begin{equation}
    \gamma_{ij} = 2 \pi \rho | \langle i | \tilde{L} |j \rangle |^2~,
\end{equation}
where $\rho$ is the density of phononic states and $\tilde{L} = \tilde{L}_{E_x}, \tilde{L}_{\epsilon_{xy}}$ is given by
\begin{equation}
    \tilde{L}= P^\dagger L P~.
\end{equation}
$P$ is a basis transformation, which allows us to restate $L$ in the eigenbasis of $H_{\rm dc}$ appearing in Eq.~\eqref{eq:time_dep_hamiltonian}. To compute $\gamma_{\rm init}$ we only need the matrix elements of $\tilde{L}$, because only the ratios between the decay rates enter Eq.~\eqref{eq:gamma init}. When numerically computing $\tilde{L}$ for $\theta_{\rm dc} = 90$ deg we find
\begin{widetext}
 \begin{align}
         \tilde{L}_{E_x} &= a (|4\rangle \langle 4|-|2\rangle \langle 2|)+b(|3\rangle \langle 3|-|1\rangle \langle 1|)+ ic(|4\rangle \langle 1|+|3\rangle \langle 2|-|1\rangle \langle 4|-|2\rangle \langle 3|),\\
        \tilde{L}_{\epsilon_{xy}} &= a'i(|1\rangle \langle 2|-|2\rangle \langle 1|+|4\rangle \langle 3|-|3\rangle \langle 4|)+b'(|1\rangle \langle 3|+|2\rangle \langle 4|+|3\rangle \langle 1|+|4\rangle \langle 2|).
 \end{align}
\end{widetext}
where $a,b \approx 10^{-1}$, $a' \approx 10^{-3}$ and $c,b' \approx 1$. From these results we conclude that $E_x$ ($\epsilon_{xy}$) phonon modes primarily couples orbital levels with opposite (same) spins, while $\epsilon_{xy}$ very weakly couples opposite spin levels of the same orbital branch.

\section{Radiative spontaneous emission}
\label{app:radiative emission}

Group theory allows us to determine the non-zero matrix elements of the transition dipole moments but not their magnitudes. This was discussed in \cite{hepp_electronic} including the relative magnitudes of the dipole moments quoted as 
\begin{align}
        &
        \begin{aligned}
            p_x = &  \begin{pmatrix}
            0 & 1 \\
            1 & 0
            \end{pmatrix} \otimes \begin{pmatrix}
             1 & 0 & 0 & 0\\
             0 & 1 & 0 & 0\\
             0 & 0 & -1 & 0\\
             0 & 0 & 0 & -1\\
        \end{pmatrix}
        \end{aligned}
        \\
        &
        \begin{aligned}
            p_y = &  \begin{pmatrix}
            0 & 1 \\
            1 & 0
            \end{pmatrix} \otimes \begin{pmatrix}
             0 & 0 & -1 & 0\\
             0 & 0 & 0 & -1\\
             -1 & 0 & 0 & 0\\
             0 & -1 & 0 & 0\\
        \end{pmatrix}
        \end{aligned} \\
        & \begin{aligned}
            p_z = &  2\begin{pmatrix}
            0 & 1 \\
            1 & 0
            \end{pmatrix} \otimes \begin{pmatrix}
             1 & 0 & 0 & 0\\
             0 & 1 & 0 & 0\\
             0 & 0 & 1 & 0\\
             0 & 0 & 0 & 1\\
        \end{pmatrix}
        \end{aligned}
    \end{align}

Based on the Wigner-Weisskopf approximation \cite{weisskopf_berechnung_1930} we find the spontaneous emission rate of G4Vs
\begin{equation}
    \gamma_{ij}= \frac{nw^3_{ij}(|\langle i|d_x|j\rangle|^2+|\langle i|d_y|j\rangle|^2+|\langle i|d_z|j\rangle|^2)}{3\pi \epsilon_0 \hbar c^3}
    \label{eq:app_fermi_gr}
\end{equation}
where $n=2.42$ is the refractive index of diamond, $\epsilon_0$ is the vacuum electric permittivity, $w_{ij}$ is the frequency splitting between the two levels $|i\rangle, |j\rangle$, and $\vec{d}=e \eta (p_x,p_y,p_z)$ is the transition dipole operator, which includes a scaling factor $\alpha$ that we seek to determine.

The total decay rate from the lowest lying excited state can be decomposed into the zero phonon line (ZPL) and phonon sideband emission rates $\gamma_{\rm total} = \gamma_{\rm ZPL} + \gamma_{\rm PSB}$. For the SnV the emission into the ZPL is given by $\gamma_{\rm ZPL} =60\% \gamma_{\rm total}$ \cite{gorlitz_spectroscopic_2020}. Given the excited state lifetime of SnV, $\tau =4.5$ns at 4 K \cite{trusheim_transform-limited_2020}, we find $\eta = 8.67635 \times 10^{-10}$. \\

\section{Rate equations}
\label{app:rate_equations}

    We calculate the rate equations for the populations in the strong pumping limit starting with the system's Hamiltonian in the rotating wave approximation, where we neglect contributions from higher lying excited states. The Hamiltonian in rotating frame is 
    \begin{equation}
        H_{\rm RWA} = 
        \begin{pmatrix}
            0 & 0 & 0 & 0 & \Omega \\
            0 & \Delta_1 & 0 & 0 & 0 \\
            0 & 0 & \Delta_2 & 0 & 0 \\
            0 & 0 & 0 & \Delta_3 & 0 \\
            \Omega & 0 & 0 & 0 & 0  \\
        \end{pmatrix}
    \end{equation}
    where $\Delta_i = E_i - \omega_L$, $E_i$ is the energy of the $i$th eigenstate $|i\rangle$ of the Hamiltonian in Eq.~\eqref{eq:ham_dc} referenced to the energetically lowest lying ground state and $\omega_L$ is the pumping laser's frequency. We only include the energetically lowest lying excited state $|5\rangle$.

    The rate equations in the quasi stationary limit generated by this Hamiltonian can be calculated from the master equation in Lindblad form:
    \begin{equation}
        \dot{\rho} = -\frac{\i}{\hbar} [H_{\rm RWA},\rho] + \sum_{\alpha} \gamma_\alpha \left(L_\alpha \rho L_\alpha^\dagger -\frac{1}{2} \{L^\dagger_\alpha L_\alpha, \rho\}\right)~, \label{eq:lindblad}
    \end{equation}
    where $\rho$ is the density matrix, the $L_\alpha \in \{L_{15},L_{25},L_{35},L_{45},L_{13},L_{23},L_{14},L_{24}\}$ and $L_{ij} = |i\rangle\langle j|$. We can calculate the optical spontaneous emission rates $\gamma_{15}, \gamma_{25}, \gamma_{35}, \gamma_{45}$ using Eq.~\eqref{eq:app_fermi_gr}. The calculation of the phononic spontaneous emission rates $\gamma_{13}, \gamma_{23}, \gamma_{14}, \gamma_{24}$ is explained in \ref{app:Phonon}.

    After evaluating Eq.~\ref{eq:lindblad}, we can find the rate equations for the quasi stationary state by assuming that $\dot{\rho}_{ij} = 0$ for $i \neq j$. We find after eliminating $\rho_{51}$ and $\rho_{15}$, using the previous assumption 
    
    \begin{widetext}
        \begin{align}
            \dot{\rho_{11}} &= \gamma_{13} \rho_{33} + \gamma_{14} \rho_{44} + \gamma_{15} \rho_{55} +\frac{4(\rho_{55}-\rho_{11})\Omega^2}{\gamma_{15}+\gamma_{25}+\gamma_{35}+\gamma_{45}}\label{eq:app_mov01}\\
            \dot{\rho_{22}} &= \gamma_{23} \rho_{33} + \gamma_{24} \rho_{44} +\gamma_{25}\rho_{55}\label{eq:app_mov02}\\
            \dot{\rho_{33}} &= -(\gamma_{23}+\gamma_{23}) \rho_{33} + \gamma_{35}\rho_{55} \label{eq:app_mov1}\\
            \dot{\rho_{44}} &= -(\gamma_{14}+\gamma_{24}) \rho_{44} + \gamma_{45}\rho_{55} \label{eq:app_mov2}\\
            \dot{\rho_{55}} &= -(\gamma_{15}+\gamma_{25}+\gamma_{35}+\gamma_{45})\rho_{55} +\frac{4(\rho_{11}-\rho_{55})\Omega^2}{\gamma_{15}+\gamma_{25}+\gamma_{35}+\gamma_{45}} \label{eq:app_mov03}    
        \end{align}
    \end{widetext}
    We further eliminate $\rho_{33}$ and $\rho_{44}$ by formally integrating Eqs.~\eqref{eq:app_mov1},\eqref{eq:app_mov2}:
    \begin{widetext}
    \begin{align}
        \rho_{33} &= e^{-(\gamma_{13} + \gamma_{23})t}\rho_{\rm 33}(0) + e^{-(\gamma_{23} + \gamma_{23})t}\int_{-\infty}^{t} e^{(\gamma_{13} + \gamma_{23})\tau} \gamma_{35} \rho_{55}(\tau) \d\tau 
        \\
        &\approx e^{-(\gamma_{13} + \gamma_{23})t}\rho_{\rm 33}(0) +  \frac{\gamma_{35}}{\gamma_{13} + \gamma_{23}} \rho_{55}~,\\
        \rho_{44} &= e^{-(\gamma_{14} + \gamma_{24})t}\rho_{\rm 44}(0) + e^{-(\gamma_{14} + \gamma_{23})t}\int_{-\infty}^{t} e^{(\gamma_{14} + \gamma_{24})\tau} \gamma_{45} \rho_{55}(\tau) \d\tau 
        \\
        &\approx e^{-(\gamma_{14} + \gamma_{24})t}\rho_{\rm 44}(0) +  \frac{\gamma_{45}}{\gamma_{14} + \gamma_{24}} \rho_{55}~,
    \end{align}
    \end{widetext}
    where we assumed that $\rho_{55}(\tau)$ has no memory and is quasi stationary. Inserting the integrated equations back into Eq.~\eqref{eq:app_mov01} allows us to integrate Eqs.~\eqref{eq:app_mov01}, \eqref{eq:app_mov02} and \eqref{eq:app_mov03}. In the limit of strong pumping we find the relevant equations for the populations shown in Eqs.\eqref{eq:pop11} and \eqref{eq:pop22}. 

\end{appendix}

\bibliography{Bib.bib}

\end{document}